\documentclass[jcp,aip,reprint]{revtex4-1}

\usepackage{graphicx}
\usepackage{dcolumn}
\usepackage{bm}

\usepackage[utf8]{inputenc}
\usepackage[T1]{fontenc}
\usepackage{mathptmx}
\usepackage{amsmath}
\usepackage{etoolbox}

\makeatletter
\def\@email#1#2{%
 \endgroup
 \patchcmd{\titleblock@produce}
 {\frontmatter@RRAPformat}
 {\frontmatter@RRAPformat{\produce@RRAP{*#1\href{mailto:#2}{#2}}}\frontmatter@RRAPformat}
 {}{}
}%
\makeatother
\begin{document}

\preprint{AIP/123-QED}

\title{Onset of Spin Entanglement in Doped Carbon Nanotubes Studied by EPR}

\author{Andreas Sperlich}
 \affiliation{Experimental Physics 6 and W\"{u}rzburg-Dresden Cluster of Excellence ct.qmat, Julius-Maximilians-Universit\"{a}t W\"{u}rzburg, Germany.}

\author{Klaus H. Eckstein}
 \affiliation{Institute of Physical and Theoretical Chemistry, Julius-Maximilians-Universit\"{a}t W\"{u}rzburg, Germany.}

\author{Florian Oberndorfer}
 \affiliation{Institute of Physical and Theoretical Chemistry, Julius-Maximilians-Universit\"{a}t W\"{u}rzburg, Germany.}

\author{Bernd K. Sturdza}
 \affiliation{Experimental Physics 6 and W\"{u}rzburg-Dresden Cluster of Excellence ct.qmat, Julius-Maximilians-Universit\"{a}t W\"{u}rzburg, Germany.}

\author{Michael Auth}
 \affiliation{Experimental Physics 6 and W\"{u}rzburg-Dresden Cluster of Excellence ct.qmat, Julius-Maximilians-Universit\"{a}t W\"{u}rzburg, Germany.}

\author{Vladimir Dyakonov}
 \affiliation{Experimental Physics 6 and W\"{u}rzburg-Dresden Cluster of Excellence ct.qmat, Julius-Maximilians-Universit\"{a}t W\"{u}rzburg,  Germany.}

\author{Roland Mitric}
 \affiliation{Institute of Physical and Theoretical Chemistry, Julius-Maximilians-Universit\"{a}t W\"{u}rzburg, Germany.}
 
 \author{Tobias Hertel}
 \affiliation{Institute of Physical and Theoretical Chemistry, Julius-Maximilians-Universit\"{a}t W\"{u}rzburg, Germany.}
 \email{tobias.hertel@uni-wuerzburg.de.}

\date{\today}

\begin{abstract}
Nanoscale semiconductors with isolated spin impurities have been touted as promising materials for their potential use at the intersection of quantum, spin, and information technologies. Electron paramagnetic resonance (EPR) studies of spins in semiconducting carbon nanotubes have overwhelmingly focused on spins more strongly localized by $\rm sp^3$-type lattice defects. However, the creation of such impurities is irreversible and requires specific reactions to generate them. Shallow charge impurities, on the other hand, are more readily and widely produced by simple redox chemistry, but have not yet been investigated for their spin properties. Here we use EPR to study p-doped (6,5) semiconducting single-wall carbon nanotubes (s-SWNTs) and elucidate the role of impurity-impurity interactions in conjunction with exchange and correlation effects for the spin behavior of this material. A quantitative comparison of the EPR signals with phenomenological modeling combined with configuration interaction electronic structure calculations of impurity pairs shows that orbital overlap, combined with exchange and correlation effects, causes the EPR signal to disappear due to spin entanglement for doping levels corresponding to impurity spacings of $14\,\rm nm$ (at 30 K). This transition is predicted to shift to higher doping levels with increasing temperature and to lower levels with increasing screening, providing an opportunity for improved spin control in doped s-SWNTs.\end{abstract}

\maketitle

\section{\label{sec1}Introduction}

The control of carrier density and spin behavior in nanoscale semiconductors is critical for enabling a wide range of device technologies.\cite{Chuang2015, Seebauer2010} Among nanoscale semiconductors, single-wall carbon nanotubes (s-SWNTs) stand out due to their atomically defined geometric structure and exceptional electronic, thermal, and mechanical properties. They are characterized by K-valley degenerate one-dimensional valence and conduction bands, tunable band gaps, and weak spin-orbit coupling, making them a compelling system for spintronic devices, among others.

Redox-chemical doping of s-SWNTs has been shown to create shallow impurity states.\cite{Eckstein2017,Eckstein2021, Eckstein2023b, Murrey2023} The correlation between doping levels and chlorine ion concentrations for gold(III)chloride doped (6,5) s-SWNTs \cite{Kim2011} has facilitated the understanding of impurity state formation mechanisms, highlighting the role of exohedrally adsorbed counterions on nanotube surfaces in generating these states. \cite{Eckstein2017} Chlorine induced impurity states are split from the valence bands and typically exhibit a binding energy of around 100 meV and an axial size of 3 to 4 nm.\cite{Eckstein2023b} The nature of the counterions, in particular their size and solvation state and thus their distance from the nanotube surface, critically affects the impurity binding energy.

Presently, EPR studies of SWNTs have primarily focused on understanding charge-transfer reactions for organic photovoltaics,\cite{Niklas2014} the effects of covalent defects,\cite{Chen1999, Corzilius2008, Zaka2010} the manipulation of spin properties for quantum applications,\cite{Trerayapiwat2021, Lohmann2020} and more recently, the potential for quantum information processing.\cite{Chen2023} Despite this broad scope, EPR spectroscopy of shallow impurities introduced by redox chemical charge transfer doping - an effective and widely used method of modifying charge transport in s-SWNTs,\cite{HermosillaPalacios2023} which is crucial for their functionality in switching devices - remains unexplored. 

\begin{figure}[htbp]
	\centering
		\includegraphics[width=8.0 cm]{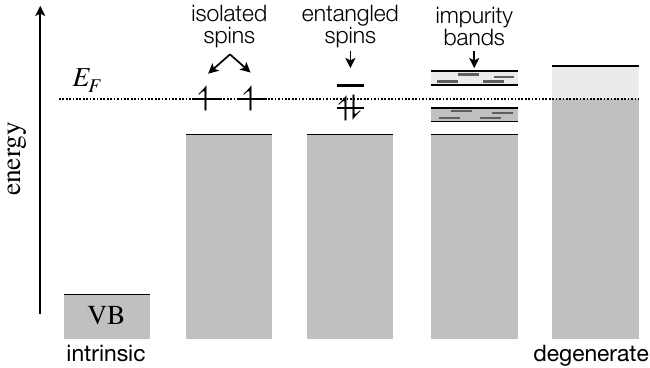}
		\caption{{\bf Progressive semiconductor doping.} With increasing doping level the semiconductor evolves from intrinsic to degenerate doping. This is initially accompanied by the formation of isolated impurity states, spin-entangled impurity pairs, impurity bands with a gap and eventually, following the insulator-metal transition, the merging of these states with the valence band (in the case of p-doping).}
		\label{fig1}
\end{figure}

This is surprising, as the interaction between shallow impurities and their effect on spin entanglement is also an important step in the transition from intrinsic to degenerately doped semiconductors (see Fig. \ref{fig1}). With increasing doping levels, the overlap of impurity wavefunctions initially leads to the formation of impurity bands. Subsequent changes in the dielectric screening contribute to further delocalisation and closing of the gap between impurity states at the insulator-to-metal or Mott transition, thus completing the transition to the degenerately doped semiconductor.

In this study we observed that the EPR signal of weakly doped (6,5) s-SWNTs disappears towards higher doping levels, similar to observations for doped semiconducting polymers.\cite{Privitera2021, Devreux1987} However, the impurity wavefunction in s-SWNTs is significantly more delocalised than in polymers, covering a few hundred carbon atoms. This raises the question if the EPR signal diminishes with increasing doping in s-SWNTs due to delocalisation or spin entanglement.

To better understand the correlation between doping levels and measured spin signals, we have used calibrated EPR data to quantitatively compare spin and charge densities. We observe that at low doping levels, each additional charge contributes a single spin to the EPR signal, suggesting that low doping levels are associated with non-interacting and singly occupied impurity levels. Measurements at 30 K show that the strength of the spin signal begins to decrease when the impurity spacings fall below about 14 nm. This behavior is modeled by electronic structure calculations of impurity pairs using configuration interactions, which allow for exchange and correlation effects on the electronic structure of interacting impurity states.

Our investigations reveal that the decay of the EPR spin signal at higher doping levels can be attributed to spin entanglement. Additionally our model suggests that temperature and relative permittivity variations influence the threshold of spin entanglement, shifting it to higher doping levels with increased temperatures and to lower levels with enhanced screening. This finding complements existing research on the impact of doping on impurity binding energies and wave function dimensions.\cite{Eckstein2023a} These insights collectively advance our understanding of the effect of doping on the electronic properties, spin interactions and spin-entanglement.

\section{\label{sec2}Methods}

The methods section covers the computational approach used for modelling impurity-impurity interactions, followed by a discussion of the experimental methodology, with emphasis on sample preparation, doping procedures as well as details on the and electron paramagnetic resonance (EPR) measurements.

\begin{figure}[htbp]
	\centering
		\includegraphics[width=8.0 cm]{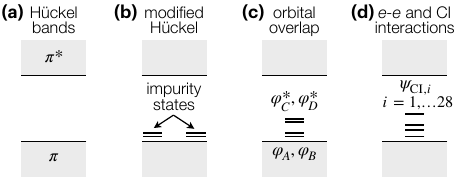}
		\caption{{\bf Illustration of model calculation hierarchy.} \textbf{(a)} H\"uckel calculations are used to obtain the valence and conduction band states of (6,5) s-SWNTs. \textbf{(b)} Modified H\"uckel calculations include interactions with exohedral counterions leading to the formation of shallow impurity states. \textbf{(c)} The effect of impurity wavefunction overlap is also calculated at the H\"uckel level with \textbf{(d)} $e-e$ interactions finally being included in the CI calculations of singly charged impurity pairs.}
		\label{fig2}
\end{figure}

The effect of spin-entangled impurity pairs for the attenuation of the EPR spin-1/2 signals is discussed by comparing a computational model with experimental results. The calculations evaluate the electronic structure of impurity pairs as a function of their spacing $s$. The latter is here inversely associated with the doping level as $n=s^{-1}$. Figure \ref{fig2} schematically illustrates the hierarchy of calculations used.

The initial H\"uckel calculations, indicated in Fig. \ref{fig2}(a), model the band structure of intrinsic (6,5) semiconducting single-wall carbon nanotubes (s-SWNTs) using a minimal atomic $p_z$ orbital basis set. The impurities induced by doping are modeled using modified H\"uckel calculations (indicated in Fig. \ref{fig2}(b)). These include the Coulomb interactions between the $\pi$ system and external chlorine counterions introduced by redox chemical doping.\cite{Eckstein2023b} Detailed discussions of the scaling behaviour, characteristics of shallow, doubly degenerate s-type impurity states with binding energies around $100\,\rm meV$ and their wavefunctions with about 2 to 4 nm in size are available elsewhere. \cite{Eckstein2023b} Such findings are consistent with experimentally observed changes in the optical spectra, including the appearance of trion bands and changes in the position, shape and intensity of exciton bands \cite{Eckstein2017,Eckstein2019, Eckstein2023a} as well as with infrared spectra.\cite{Eckstein2021}

The interactions between impurity states are here analysed within the modified H\"uckel framework.\cite{Eckstein2023b} The formation of bonding and anti-bonding states from the overlap of impurity wavefunctions is similar to those observed in the molecular hydrogen ion, as indicated in Fig. \ref{fig2}(c). To include carrier-carrier interactions, specifically for hole states, each impurity is assumed to carry a single charge. We then diagonalize the configuration interaction (CI) Hamiltonian with respect to twenty eight two particle spin orbitals resulting from the four H\"uckel orbital wavefunctions and the four two-electron spin configurations. This approach yields the energy splitting of the two-particle state energies as the impurity spacings decrease, falling into four groups of states as shown schematically in Fig. \ref{fig2}(d).

\subsection{\label{subsec2.1} Modified H\"{u}ckel calculations}

In this study, a previously introduced modified H\"{u}ckel model is used to simulate the $\pi$ system of the nanotube using a minimal $p_z$ orbital basis set. We give a brief overview of these calculations, with more details available in a previous publication.\cite{Eckstein2023b} The versatility of this basis set to describe even complex phenomena in s-SWNTs, such as exciton quasi-particle energies, has been demonstrated by Perebeinos et al.\cite{Perebeinos2004}

The modified H\"{u}ckel calculations include Coulomb interactions between $\pi$ electrons and one or more exohedral counterions. To this end we have modified the H\"{u}ckel Coulomb integrals, $\alpha_i$, accordingly. Corrections are also applied to the nearest neighbour resonance integrals, $\gamma$, to reflect curvature induced changes as discussed by Ding et al.\cite{Ding2002,Hagen2003} This correction becomes increasingly important for smaller diameter SWNTs.

For best agreement with the empirical band gaps of (6,5) s-SWNTs, of approximately 1.5 eV,\cite{Eckstein2017} and with the predicted free carrier absorption thresholds,\cite{Pedersen2004, Perebeinos2004} a nearest neighbour resonance integral $\gamma_0$ of 4 eV was chosen. In this study, we calculate all orbital energies with respect to the center of the band gap.

The (6,5) nanotubes studied here have a diameter of 757 pm. Their computational model geometry contains 364 carbon atoms per unit cell. To simulate low doping levels, corresponding to carrier concentrations as low as $0.03\,{\rm nm}^{-1}$, 10 unit cells were used, representing a total of 5,460 carbon atoms along a 62 nm nanotube segment. Periodic boundary conditions were used to construct the H\"{u}ckel matrix for this structure.

These calculations were carried out for impurity pairs at various axial separations $s$, yielding two bonding and two anti-bonding doubly degenerate impurity pair orbitals, denoted $\varphi_\mu$ and $\varphi_\mu^*$ respectively. The wavefunctions of these four orbitals are similar to the bonding and anti-bonding wavefunctions observed in $H_2^+$, with the anti-bonding wavefunctions having a node centered between the two impurity sites.

\subsection{\label{subsec2.2} The two particle CI problem}

To account for carrier-carrier and exchange interactions, we next considered a very simple configuration interaction (CI) problem for two electrons distributed over $N_{\rm orb}$ spatial orbitals. Note that the two-electron problem is homologous to the two-hole problem faced for the impurity pairs of \textit{p}-doped nanotubes. By combining $N_{\rm orb}=4$  spatial orbitals with two spin states we can now construct a set of $2\times N_{\rm orb}$ spin-orbitals:
\begin{equation}
    \{\phi_i, i=0,\cdots 2\times N_{\rm orb}-1 \}.
    \label{teq1}
\end{equation}
Each spin orbital $\phi_i$ is defined as the product of a spatial part $\varphi_\mu$ where $\mu$ ranges from 0 to $N_{\rm orb}-1$ and the spin index $\sigma$ representing two possible spin down and up states. A general spin orbital can thus be written as
\begin{equation}
    \phi_i=\varphi_{\mu(i)}\,\sigma(i).
    \label{teq2}
\end{equation}
From the above set of $N_{\rm orb}$ spin-orbitals we can thus build total of $\tfrac{1}{2}\times (2N_{\rm orb})\times (2N_{\rm orb}-1)$ distinct configuration state functions containing two electrons. Each such configuration is defined by specifying two spin-orbital indices $i$ and $j$ and has the form of a normalized Slater determinant:
\begin{equation}
    |ij\rangle = \frac{1}{\sqrt{2}} \begin{vmatrix}
    \phi_i(1) & \phi_j(1) \\
    \phi_i(2) & \phi_j(2) \\
    \end{vmatrix} = \frac{1}{\sqrt{2}}[\phi_i(1)\phi_j(2)-\phi_i(2)\phi_j(1)]
    \label{teq3}.
\end{equation}

\subsubsection{The Hamiltonian and the CI-expansion}

To model interactions between neighboring impurities we then use the following two-electron effective Hamiltonian:
\begin{equation}
    \hat H=h(1)+h(2)+V_{\rm Ohno}(r_{12}) 
    \label{teq4}.
\end{equation}
Here, $h$ is the one electron, modified H\"uckel Hamiltonian described in the previous sections and in Ref. \onlinecite{Eckstein2023b}. The term $V_{\rm Ohno}(r_{12})$ is the Ohno potential modeling Coulomb interactions between different lattice sites (in atomic units), 
\begin{equation}
    V_{\rm Ohno}(r_{12})=\frac{U}{\sqrt{(U\epsilon_r \, r_{12})^2+1}}
    \label{teq4b},
\end{equation}
where $\epsilon_r$ is the relative permittivity and $U=0.415\,\rm a.u.\equiv 11.3\,\rm eV $ is the Hubbard parameter describing the energy cost of placing two electrons on the same site. This approach describes organic polymer systems well and has previously been used to study the electronic and optical properties of carbon nanotubes \cite{Perebeinos2004} and other organic polymer systems.\cite{Heeger1988,Abe1992}

The configuration interaction (CI) wavefunction is given as a linear combination of all Slater-determinants that can be constructed within the restricted active space containing $2N_{\rm orb}$ spin-orbitals:
\begin{equation}
    \left|\psi_{\rm CI}\right\rangle =\sum_i\sum_{j<i}C_{ij}|ij\rangle
    \label{teq5}.
\end{equation}
Since our active space contains 8 spin-orbitals the CI-expansion contains 28 possible configurations (Slater determinants) that can be separated into 10 singlet and 6 triplet spin-eigenstates. In our approach we do not perform spin-adaptation in advance but diagonalize the full CI-Hamiltonian first and then identify the spin eigenstates by analyzing the CI eigenstates. 

\subsubsection{Construction of the CI matrix}
The CI states can be calculated in a standard manner by solving the following matrix eigenvalue problem:
\begin{equation}
    \sum_{ij}H_{kl,ij}\,C_{ij}=E\, C_{ij}\quad \forall k=0,2N_{\rm orb}-1;l=k,2N_{\rm orb}-1
    \label{teq6}.
\end{equation}
The CI -Matrix $H_{kl,ij}$ is defined as the matrix element of the Hamiltonian between two configurations $|ij\rangle$ and $|kl\rangle$ :
\begin{equation}
    H_{kl,ij}=\langle ij|\hat H|kl\rangle 
    \label{teq7},
\end{equation}
that can be easily evaluated using standard manipulations. We here provide the expressions for completeness:
\begin{align}
    H_{kl,ij}&=\langle ij|h(1)+h(2)+V_{\rm Ohno}|kl\rangle \nonumber\\
    &=\langle ij|h(1)|kl\rangle+\langle ij|h(2)|kl\rangle+\langle ij|V_{\rm Ohno}|kl\rangle \nonumber\\
    &=2\langle ij|h(1)|kl\rangle+\langle ij|V_{\rm Ohno}|kl\rangle
    \label{teq8}.
\end{align}
We use $h=h(1)$ such that the one-electron part becomes:
\begin{align}
    \langle ij|h|kl\rangle &= \frac{1}{2}[ \langle \phi_i|h|\phi_k\rangle \delta_{jl} - \langle \phi_i|h|\phi_l\rangle \delta_{jk} \nonumber \\
    & - \langle \phi_j|h|\phi_k\rangle \delta_{il} + \langle \phi_j|h|\phi_l\rangle \delta_{ik}]
    \label{teq9}.
\end{align}
For the two-electron part we have
\begin{align}
    \langle ij|V_{\rm Ohno}|kl\rangle &=  \langle \phi_i\phi_j|V_{\rm Ohno}|\phi_k\phi_l\rangle - \langle \phi_i\phi_j|V_{\rm Ohno}|\phi_l\phi_k\rangle 
    \label{teq10}.
\end{align}
By joining the one- and two-electron parts we obtain a general expression for the matrix elements expressed in terms of spin-
orbitals:
\begin{align}
    H_{ij,kl}&=\langle\phi_i|h|\phi_k\rangle\delta_{jl}-\langle\phi_i|h|\phi_l\rangle\delta_{jk}\nonumber \\
    &-\langle\phi_j|h|\phi_k\rangle\delta_{il}+\langle\phi_j|h|\phi_l\rangle\delta_{ik}\nonumber \\
    &+\langle \phi_i\phi_j|V_{\rm Ohno}|\phi_k\phi_l\rangle - \langle \phi_i\phi_j|V_{\rm Ohno}|\phi_l\phi_k\rangle
    \label{teq11}.
\end{align}
Now we can separate the spin and the spatial part. To do this we define functions $\mu(i)$ and $\sigma(i)$ that associate spatial and spin indices to each spin orbital. This leads to:
\begin{align}
    H_{ij,kl}&=\langle\varphi_{\mu(i)}|h|\varphi_{\mu(k)}\rangle \delta_{\sigma(i)\sigma(k)}\delta_{jl}\nonumber\\
    &-\langle\varphi_{\mu(i)}|h|\varphi_{\mu(l)}\rangle\delta_{\sigma(i)\sigma(l)}\delta_{jk}\nonumber\\
    &-\langle\varphi_{\mu(j)}|h|\varphi_{\mu(k)}\rangle\delta_{\sigma(j)\sigma(k)}\delta_{il}\nonumber\\
    &+ \langle\varphi_{\mu(j)}|h|\varphi_{\mu(l)}\rangle\delta_{\sigma(j)\sigma(l)}\delta_{ik}\nonumber\\
    &+\langle\varphi_{\mu(i)}\varphi_{\mu(j)}|V_{\rm Ohno}|\varphi_{\mu(k)}\varphi_{\mu(l)}\rangle\delta_{\sigma(i)\sigma(k)}\delta_{\sigma(j)\sigma(l)}\nonumber \\
    &+\langle\varphi_{\mu(i)}\varphi_{\mu(j)}|V_{\rm Ohno}|\varphi_{\mu(l)}\varphi_{\mu(k)}\rangle\delta_{\sigma(i)\sigma(l)}\delta_{\sigma(j)\sigma(k)}
    \label{teq12}.
\end{align}
This is the working expression for the CI matrix that can be easily implemented into a computer code. In this expression the two-electron integrals appear including the spatial parts of the spin orbitals. Since each spatial orbital can be expanded into the atom-centred basis functions (in our case carbon p-orbitals) according to
\begin{equation}
    \varphi_\mu=\sum_i c_i^\mu p_i 
    \label{teq13},
\end{equation}
the two-electron integrals over MOs can be expressed in terms of atomic two-electron integrals involving the four index transformation:
\begin{equation}
    \langle\varphi_{\mu}\varphi_{\nu}|V_{\rm Ohno}|\varphi_{\lambda}\varphi_{\sigma}\rangle=\sum_{ijkl}c_i^\mu c_j^\nu c_k^\lambda c_l^\sigma \langle p_ip_j|V_{\rm Ohno}|p_kp_l\rangle
    \label{teq14}.
\end{equation}

Since the atomic orbitals are localized, we can use the "zero differential overlap" approximation to simplify the above expression. Assuming that:
\begin{align}
    p_i(1)p_k(1)&=|p_i|^2\delta_{ik}\nonumber \\ 
    p_j(2)p_l(2)&=|p_j|^2\delta_{jl} 
    \label{teq15},
\end{align}
we obtain
\begin{align}
    \langle\varphi_{\mu}\varphi_{\nu}|V_{\rm Ohno}|\varphi_{\lambda}\varphi_{\sigma}\rangle &=\nonumber \\
    \quad \sum_{ij}&c_i^\mu c_j^\nu c_i^\lambda c_j^\sigma \iint d1\,d2\, V_{\rm Ohno}|p_i(1)|^2|p_j(2)|^2
    \label{teq16}.
\end{align}
The on-site integrals ($i = j$) are replaced by the Hubbard-parameter $U_i$ and the pair-terms can be approximated by the Ohno potential leading to:
\begin{align}
    \sum_{ij}& c_i^\mu c_j^\nu c_i^\lambda c_j^\sigma\, U_i + \sum_i\sum_{j \neq i}V_{\rm Ohno}c_i^\mu c_j^\nu c_i^\lambda c_j^\sigma
    \label{teq17}.
\end{align}

\subsection{\label{subsec2.3} Experiment}

\subsubsection{\label{subsec2.3.1} Preparation, concentration and doping of nanotube samples}

Sample preparation for the study of chirality-enriched single-wall carbon nanotubes (SWNTs) begins with the dispersion of carbon-rich nanotube soot in an organic solvent, typically toluene, using polymers as dispersion additives, a method pioneered by Nish et al. \cite{Nish2007} and further refined by Ozawa et al. for (6,5)-SWNTs.\cite{Ozawa2011} This refined process includes shear mixing to minimize structural defects, employing techniques similar to those described by Graf et al.\cite{Graf2016}

The starting material, raw CoMoCat nanotube soot (SG65i, Sigma Aldrich), was added to a solvent mixture with a 5:1 volume ratio of toluene to acetonitrile, at a typical carbon content of about $1\rm\, mg/ml$. This mixture was agitated for 15 minutes using a conventional bath sonicator to initiate disentanglement of the nanotube fibers and aggregates. Subsequently, a nearly equivalent mass per volume of matrix polymer, [(9,9-dioctylfluorenyl-2,7-diyl)-alt-co-(6,6'-2,2'-bipyridine)], abbreviated as PFO-BPy, was added, and the mixture underwent another 15 minutes of bath sonication.

Following this initial treatment, the suspensions were further dispersed at $20^\circ\mathrm{C}$ using shear-mixing for up to 4 days, minimizing sonication-induced damage to the nanotubes. Benchtop centrifugation then separated individualized nanotubes from aggregates with higher sedimentation coefficients. This yielded a supernatant highly enriched in (6,5)-SWNT species wrapped with PFO-BPy polymer, effectively eliminating other chiralities and impurities. This polymer-sorted SWNT material was used for further processing.

For EPR measurements, the concentrations of the nanotube suspensions required further concentration. To this end, vacuum filtration was employed using $0.1 \rm \mu m$ Millipore Membrane Filters with a standard Millipore Microanalysis filtration setup. This process retained the majority of the dispersed SWNT material as a cake on the filter for further processing.

The nanotubes were then redispersed by first transferring the filter membrane with its cake into an acetone bath. This was agitated for 20 minutes using an IKA orbital minishaker at 1 Hz. The acetone was then replaced with fresh solvent and this process was repeated three times to completely dissolve and remove all the filter material. The free-floating SWNT film was then placed in a fresh acetone bath overnight. Finally, it was redispersed in the desired amount of a 5:1 volume ratio toluene/acetonitrile solvent mixture using a bath sonicator for 45 minutes at a temperature of $20^\circ \rm C$ (Bandelin Sonorex Super 10 P).

The doping of (6,5)-SWNT:PFO-BPy samples was performed according to a procedure introduced by Kim et al.,\cite{Kim2008} where appropriate aliquots of $\mathrm{AuCl_3}$ solution are titrated into the nanotube polymer suspension. For EPR measurement series at different doping levels, the EPR sample cell was evacuated and purged with helium several times to remove residual air and moisture as much as possible without evaporating the solvent. Directly before data acquisition, the sample was sonicated at $50^\circ\mathrm{C}$ for 15 minutes.

After sonication, the sample was placed in an EPR setup and cooled to $30\,\rm K$ for spin counting measurements. To gradually increase the doping level, the sample was heated to $298\,\rm K$ before repeating the doping process by titrating additional amounts of $\rm AuCl_3$ solution. The series of EPR spectra were then used to distinguish, identify and quantify initial atmospheric and the chemically-induced \textit{p}-doping in the (6,5) s-SWNTs.

\subsubsection{\label{subsec2.3.2} EPR data acquisition and calibration}

A modified 9.4 GHz X-band spectrometer (Bruker E300) was used to collect the EPR spectra. The sample was placed in a resonant cavity (Bruker ER4104OR) and cooled with a continuous flow helium cryostat insert (Oxford ESR900). The g-factor was calibrated using an NMR Gaussmeter (Bruker ER035M) and a microwave frequency counter. EPR was measured using lock-in detection (Ametek 7230DSP) with a 100 kHz modulation of the external magnetic field as a reference. Unless stated otherwise, spectra were recorded at T = 30 K, 0.2 mW microwave power (30 dB attenuation setting) and 1 G magnetic field modulation. These parameters are necessary to operate the setup in the linear response regime and to avoid spectral saturation. \cite{Poole1996, Yordanov1994}.

For quantitative EPR measurements, calibration was achieved using a spin reference sample, specifically 1,3-bis(diphenylene)-2-phenylallyl (BDPA), which closely resembles defects and localized spins in organic materials. Its carbon-centered radical nature and propeller-like geometry provide steric protection, enhancing its chemical stability. The unpaired electron predominantly resides at the allyl core's 1- and 3-positions, with additional stabilization through delocalization over the attached biphenyl rings.\cite{Azuma1994} The reference sample, a 1:1 crystalline complex of BDPA with benzene ($\rm C_{33}H_{21} \cdot C_{6}H_{6}$), has a molar mass of 495.63 g/mol. Precise control over the material amount in the reference sample allows for accurate determination of the spin count contributing to the EPR signal. Reference samples contained $1.0 \pm 0.1,\mu g$ of BDPA, corresponding to $(1.22 \pm 0.12) \times 10^{15}$ molecules.The method is comprehensively discussed by Eaton et al. \cite{Eaton2010} 

To compare the spin densities with the number of carbon atoms in the sample, we used published optical absorption cross sections of the first subband exciton transition.\cite{Schoppler2011, Streit2014} The doping level was also determined from the optical absorption spectra using the reduction in exciton oscillator strength as a calibrated measure of the number of charges introduced by doping.\cite{Eckstein2019}

\subsection{\label{subsec2.4} Miscelaneous}
The manuscript text was edited with the assistance of OpenAI's ChatGPT-4 turbo version, accessed via its browser-based command-line interface. The primary objective was to solicit the LLM's assistance in proofreading and iteratively refining human-generated text. A typical prompt to the LLM might read, '\textit{Please proofread, check for clarity, flow, and redundancies: MANUSCRIPT TEXT}'. Subsequently, the texts were iterated by using the 'DeepL SE Write' (beta) interface (deepl.com) to preserve the original intent and ideas of their human authors.

\section{\label{sec3}Results and Discussion}

\subsection{\label{subsec3:1}EPR spectra from intrinsic and doped s-SWNTs}

\begin{figure}
    \includegraphics[width=8.0 cm]{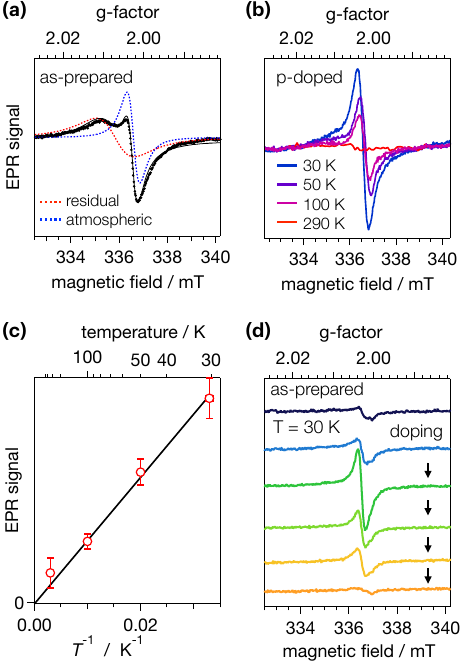}
    \caption{\textbf{EPR spectra of p-doped SWNTs}  \textbf{(a)} EPR spectrum of an as-produced nanotube sample showing two signal contributions from residual spin impurities (red dashed line) and impurities from atmospheric p-doping (blue dashed line). \textbf{(b)} Temperature dependence of EPR spectra from a moderately doped (6,5) s-SWNT sample. \textbf{(c)} Temperature dependence of the spin 1/2 signal compared with the expected Curie behavior illustrating the paramagnetic character of spin-impurities at small carrier concentrations. \textbf{(d)} Series of EPR spectra measured at 30 K from as-prepared and an increasingly (top to bottom) redox-chemically, p-doped (6,5)-SWNT sample.}
    \label{fig3}
\end{figure}

A representative EPR spectrum of an as-prepared (6,5) s-SWNT:PFO-BPy sample at 30 K is shown in Figure~\ref{fig3}(a). The signal consists of two distinct Lorentzian-shaped contributions. First, we can assess that both signals are indeed from SWNTs, since the PFO-BPy wrapping polymer alone has a residual radical concentration of less than $0.20(4) \times 10^{-5}$ spins per carbon atom of the polymer, independent of $\rm AuCl_3$ addition and sonication; i.e. more than an order of magnitude less spin concentration than the two SWNT signals. 

The first SWNT signal has a Land\'e g-factor of 2.0069(2) and a peak-to-peak line width (lwpp) of 0.70 mT. This is significantly broader than EPR signals with a g-factor of 2.0037 found for $\rm sp^3$-type defects induced by diazonium-assisted covalent functionalization of (6,5) nanotubes.\cite{Lohmann2020} The signal strength corresponds to about $6.3(8) \times 10^{-5}$ spins per carbon atom. For the (6,5) s-SWNT with 88 carbon atoms per nm, this corresponds to one such residual spin impurity per approximately 200 nm of nanotube length. Given the pronounced electronic perturbations introduced, for example, by Stone-Wales defects, together with other potential non-$\rm sp^3$-type structural anomalies, it is plausible that these could underlie the unique EPR features observed at a g-factor of 2.0069. This would be consistent with the complex interplay between the structural integrity and electronic properties of nanotubes, but would need to be verified by further detailed calculations and experiments.

The second EPR component in Figure~\ref{fig3}(a) appears with a lower g-factor of 2.00235(10), here with an lwpp of 0.26 mT. This is very close to the free electron with its g-factor of 2.00232. This lower g-factor component is attributed to residual atmospheric p-doping at a level of $4.6(5) \times 10^{-5}$ spins per carbon atom, or equivalently one dopant impurity per 250 nm of nanotube length. Such p-doping has previously been attributed to redox chemical doping from residual water and oxygen.\cite{Aguirre2009}

In Fig. \ref{fig3}(b), we present a series of spectra at different temperatures from a $\rm AuCl_3$ doped s-SWNT sample. The g-factor measured is 2.00284(10), which is only marginally higher than that of the signal component attributed to atmospheric doping. Notably, the latter appears unaffected by the redox reaction with $\rm AuCl_3$. The temperature dependence of the p-doping induced EPR band, as shown in Fig.~\ref{fig3}(c), clearly follows Curie’s law, confirming the paramagnetic nature of this spin signal.

In Fig. \ref{fig3}(d) we present a series of spectra covering a wide range of doping levels, from intrinsic to degenerately doped. This increase in doping level up to about $0.3\,\rm nm^{-1}$ is determined by absorption spectroscopy from the bleaching of the first subband exciton ($X_1$).\cite{Eckstein2019} As the doping level increases, the EPR signal initially intensifies, reaching its peak at a hole density of about $0.07\,\rm nm^{-1}$, before decreasing at higher doping levels and finally vanishing at the highest doping concentrations studied. Throughout this range of doping levels, both the position and width of the EPR signal remain essentially unchanged, indicating no significant variation in the character of the charged impurities.

No signal corresponding to $\rm AuCl_3$ was observed in the spectral range discussed above. However, in highly concentrated $\rm AuCl_3$ solutions an additional, very broad EPR signal can be detected at g values between 2.0 and 2.2 (not shown), which we tentatively attribute to multinuclear Au(II) complexes.

\subsection{\label{subsec3:2}Character of impurity states}

\begin{figure}
    \includegraphics[width=8.0 cm]{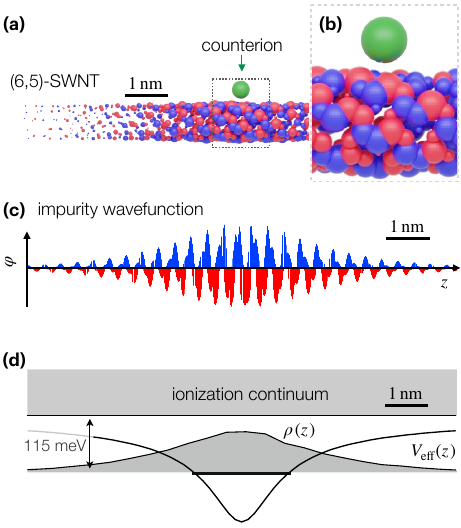}
    \caption{\textbf{Counterion induced shallow impurities.} \textbf{(a)} Calculated nanotube impurity orbital as induced by interaction with a nearby chlorine counterion (green sphere) and \textbf{(b)} magnification of the wavefunction in direct proximity of the counterion. Spheres corresponding to negative atomic orbital coefficients are colored red while positive contributions are colored blue. The sphere volume represents the orbital contribution to the charge density. \textbf{(c)} The coefficients of the impurity wavefunction plotted on the same x-scale as in d along the nanotube axis. The period of oscillations between positive and negative contributions corresponds to the axial momentum component at the top of the valence band of the unperturbed nanotube. \textbf{(d)} Schematic representation of the effective one dimensional Coulomb potential $V_{\rm eff}$ leading to the energetic alignment and spatial distribution of the impurity wavefunction.}
    \label{fig4}
\end{figure}

Doping in (6,5) s-SWNTs at low levels up to about $0.1\,\rm nm^{-1}$ has been associated with the formation of localized, shallow impurity states. This is consistent with observed changes in exciton band strength, shape, and position,\cite{Eckstein2019, Eckstein2023a} with ultrafast exciton dynamics,\cite{Eckstein2017} with IR spectral responses,\cite{Eckstein2021} as well as with electrical transport studies \cite{Murrey2023} of similarly redox-doped s-SWNTs. The character of these shallow impurity states has been studied using simple quantum chemical methods, in particular the modified H\"uckel approach described in the methods section of this paper.\cite{Eckstein2023b} Their wavefunctions are found to be completely circumferentially delocalized, extending along the axial direction of the s-SWNTs by 2 to 4 nm. This depends on the counterion type, its distance from the tube and the screening by the environment.\cite{Eckstein2023b} The character of the wavefunction can be seen from the schematic illustration of the impurity wavefunction in Fig. \ref{fig4}, including the coefficient weights for the $p_z$ orbitals of over 250 C atoms.

The localized nature of these impurity states has been attributed to the role of counterions in the doping mechanism \cite{Kim2011,Eckstein2023b} similar to the localization of the $\pi$ system wavefunctions in semiconducting thiophene (P3HT) polymers induced by external charges reported by Niklas et al.\cite{Niklas2013}  As with any redox chemical process, such counterions are abundant in the solvent environment and are likely to be bound to the doped nanotubes when these acquire charges from the redox reaction. For the $\rm AuCl_3$ hole doping reaction used here, chlorine is thought to act as an impurity stabilising counterion as illustrated schematically in Fig. \ref{fig4}(a) and (b). Under the conditions of our experiments and using an effective relative solvent permittivity of 9.25, we estimate the Coulomb interaction with the counterion to support a doubly degenerate impurity state $\varphi$ with a binding energy of about $115\,\rm meV$ and a FWHM of the charge density envelope shown in Fig. \ref{fig4}(d) of $3.6\,\rm nm$.\cite{Eckstein2023b}

The above impurity wavefunctions and binding energies are obtained for what are known in electrochemical interface terms as specifically adsorbed ions in the Helmholtz layer. Non-specifically adsorbed ions in the Guy-Chapman-Stern model of double layers are located in the so-called outer Helmholtz plane, which adds the thickness of the ion solvation shell to the ion-surface distance. Thus, if we allow for the possibility of non-specific counterion adsorption by conservatively adding 360 pm of solvation shell thickness to the counterion-tube distance (as a likely upper bound), these shallow states become slightly less localized with an FWHM of about 4.1 nm. Accordingly, the impurity is then only bound by 90 meV. In the later discussion of the EPR signal we will consider both scenarios leading to the formation of impurity states: interaction with directly adsorbed counterions ("bare") and interaction with solvated counterions ("solvated").

The observation of isolated spins by EPR in these samples is thus consistent with the presence of singly occupied, shallow and localized impurity states, as the localization of impurity wavefunctions is crucial for the detection of EPR signals from isolated spins. In addition, the binding energy of around 100 meV effectively binds charges, in this case holes, to the impurity, especially at low temperatures, further supporting the observed EPR phenomena.

\subsection{\label{subsec3:3}Character and spin of impurity pairs}

\begin{figure}
    \includegraphics[width=8.0 cm]{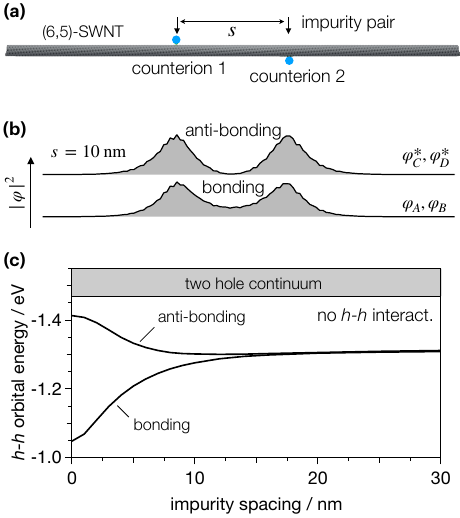}
    \caption{\textbf{Character of impurity pairs excluding \textit{h-h} interactions} \textbf{(a)} Schematic representation of the carbon nanotube with two adsorbed counterions spaced by \textit{s}. \textbf{(b)} Anti-bonding and bonding orbital wavefunctions of an impurity pair excluding hole-hole interactions. \textbf{(c)} Dependence of the corresponding orbital energies on the impurity spacing.}
    \label{fig5}
\end{figure}

At higher doping levels and hence smaller impurity pair spacings $s$ (Fig. \ref{fig5}(a), the wavefunctions of neighbouring sites are expected to overlap, leading to the formation of characteristic bonding and anti-bonding states. This can be seen in Fig. \ref{fig5}(b), where we show the charge density envelopes of the impurity pair problem using the modified H\"uckel approach. Here the impurities are spaced $s=$10 nm apart. The two degenerate bonding orbitals in Fig. \ref{fig5}(b) are labelled $\varphi_A,\varphi_B$, while the anti-bonding molecular orbitals (MOs) are labelled $\varphi_C^*,\varphi_D^*$. 

For large values of $s$ the four orbital energies and charge density envelopes converge asymptotically to those of two independent impurities. Using these orbital energies as the basis for two-hole wavefunctions, while excluding $h-h$ interactions, yields a two-hole configuration of bonding and anti-bonding MOs whose energy dependence on impurity spacing is shown in Fig. \ref{fig5}(c). Note that the orbital energies here are chosen to decrease in the upward direction to allow a discussion in the more familiar form when two electron systems are considered instead. The two-hole continuum corresponds to the orbital energy of two holes in the valence band continuum.
\begin{figure}
    \includegraphics[width=8.0 cm]{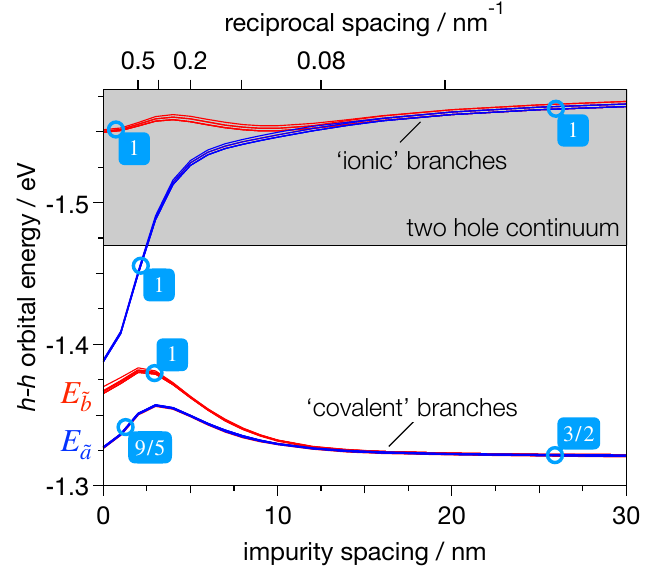}
    \caption{\textbf{Orbital energies of the CI Hamiltonian as a function of impurity distance.} At large impurity spacings, the electronic energies of the 28 eigenfunctions fall into two groups, one of covalent character and one of ionic character. At smaller impurity spacings both branches split again into a total of four branches each with a different spin expectation value $\langle \hat S^2\rangle$ (in units of $\hbar^2$ as indicated by the numbers in the blue rectangles). The splitting of the lower two branches  $\Delta E_{\tilde b\tilde a}=E_{\tilde b}-E_{\tilde a}$ and its relation to thermal energies is used to determine the spacing at which the two impurity spins become entangled.}
    \label{fig6}
\end{figure}

To account for carrier-carrier interactions as well as exchange and correlation effects, we next performed configuration interaction (CI) calculations using the screened Ohno potential, as described in the Methods section. The two bonding and anti-bonding H\"uckel orbitals are used to construct the antisymmetrized spin-orbital basis, which consists of a total of $\tfrac{1}{2}\times (2N_{\rm orb})\times (2N_{\rm orb}-1)=28$ different two-particle wave functions ($N_{\rm orb.}=4$):
\begin{equation}
 |ij\rangle=\tfrac{1}{\sqrt{2}}\left[\phi_i(1)\phi_j(2)-\phi_i(2)\phi_j(1)\right].
 \label{eq1}
\end{equation}
Here the $\phi_i=\varphi_{\mu(i)}\alpha(i)$ represent non-symmetrized spin orbitals with the $\mu(i)$ index pointing to one of the four bonding or anti-bonding modified H\"uckel wavefunctions $\varphi_A,\varphi_B,\varphi_C^*$ or $\varphi_D^*$ and $\alpha(i)$ representing one of the two possible spin orientations. This basis is diagonalized for the Hamiltonian (in atomic units)
\begin{equation}
 \hat H=h(1)+h(2)+V_{\rm Ohno}(r_{12})+r^{-1},
 \label{eq2}
\end{equation}
where $h(1)$ and $h(2)$ are the independent particle Hamiltonians of charges 1 and 2, while the Coulomb interaction between $p_z$ orbitals is modeled by the Ohno potential with an on-site Hubbard term of 11.3 eV and -- for computational simplicity -- a single dielectric constant $\epsilon_r = 4$ as used in Ref. \onlinecite{Perebeinos2004}. Note, in using a relatively low dielectric constant we here follow the established treatment of carrier-carrier interactions in $\pi$ systems whereas the treatment of the interaction of the exohedral counterion with the electronic system of the nanotube is charcterized by a higher dielectric constant. The latter is based on the low frequency dielectric response of the solvent due to molecular rearrangements in the environment of the counterion. This appears to be a sensible approach given the enormous complexities of more rigorous treatments. Finally, the counterion repulsion - as represented by the $r^{-1}$ term in eq. \ref{eq2} - can be separated from the electronic Hamiltonian.

The dependence of the electronic part of the resulting two-spin CI eigenenergies on the impurity spacing is shown in Fig. \ref{fig6}. At large impurity spacings the eigenstates fall into two groups, with the sixteen lowest states representing what are commonly referred to as 'covalent' states, where the carrier density is evenly distributed between the two impurity sites. The twelve states of the upper branch are representative of 'ionic' configurations which are offset to the covalent branch by the Coulombic energy cost of confining both carriers to a single impurity.

This splitting between covalent and ionic states is essentially what gives rise to the gap between impurity bands at intermediate doping levels and, depending on the doping level, the 'insulating' state of doped semiconductors. Interestingly, due to the small impurity binding energy, the covalent-ionic splitting of about $250\,\rm meV$ pushes the upper branch of states above the ionisation threshold, suggesting that such states are likely to couple with the delocalized states of the ionisation continuum.

At impurity spacings below about 10 nm, the CI states fall into 4 groups. The lowest energy CI group with 10 spin states is composed of three triplet-states ($S_i=1$ for $i=1,3,4$) and one singlet-state ($S_2=0$). The average energy of states in this group is designated as $E_{\tilde a}$. The higher energy covalent branch with 6 spin states is composed of one triplet- ($S_5=1$) and three singlet-states ($S_i=0$ for $i=6,7,8$) and its average energy is given by $E_{\tilde b}$. The indices are ordered to reflect increasing energy of the respective spin orbital at small impurity spacings. The lower of the ionic branches consists of one triplet-state ($S_6=1$) and three singlet-states ($S_i=0$ for $i=7,8,9$) while the highest energy ionic branch with 6 spin states is similarly composed of one triplet-state ($S_10=1$) and three singlet-states ($S_i=0$ for $i=11,12,13$). 

The spin expectation value of each of these branches is given by $\langle \hat S^2\rangle =N^{-1}\sum (2S_i+1) \hat S_i^2$, where $N$ is the total number of spin states within a branch. With spin eigenvalues given by $\hat S_i^2=S_i(S_i+1)\hbar^2$ and $N=10$ for the lowest energy branch at small impurity spacings, we then obtain the expectation value of $\tfrac{9}{5}\,\hbar^2=1.8 \hbar^2$ (see Fig. \ref{fig6}). Expectation values for the other groups of CI states in the covalent and ionic branches are also shown in Fig. \ref{fig6} in units of $\hbar^2$. At large impurity spacings the covalent branch has a lower spin expectation value of only $3/2 \hbar^2$. This change in expectation value indicates that energetic splitting at spacings close to 10 nm may lead to a change in the experimentally observed EPR signals.

\begin{figure}
    \includegraphics[width=8.0 cm]{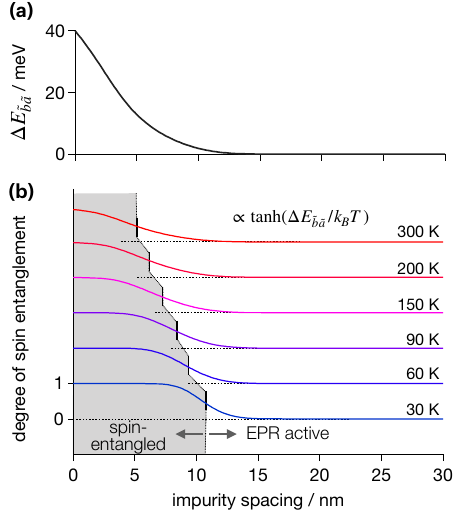}
    \caption{\textbf{CI branch energy splitting and spin entanglement.} \textbf{(a)} The energy splitting $\Delta E_{\tilde b\tilde a}$ between the lowest CI branches increases to about 40 meV at the smallest impurity spacings. \textbf{(b)} Waterfall plot with the impurity spacings indicated at which thermal excitations become greater than $\Delta E_{\tilde b\tilde a}$ decreases with increasing temperature. This implies that spin entanglement at higher temperatures requires greater overlap between impurity wavefunctions.}
    \label{fig7}
\end{figure}

At the same time it is important to remember that at large impurity spacings, where the spins no longer interact, the actual spin expectation will be that of a thermal ensemble of independent spin 1/2 states with $\frac{3}{2}\hbar^2$, where each of the two spins contributes $\frac{3}{4}\hbar^2$. Whereas in the CI model the spin expectation values of the ionic and covalent branches are imputed by virtue of the Slater determinant Ansatz used for the CI calculations. The latter artificially imposes spin entanglement independent of impurity spacing or system temperature.

\subsection{\label{subsec3:4}Intensity of the EPR spin 1/2 signal}

In order to model the disappearance of the EPR spin signal in the doped nanotubes, it is necessary to find a sensible description of the onset of the formation of entangled spin states in the impurity pair. However, as mentioned above, the CI calculations imply spin entanglement from the outset at any temperature and impurity spacing, clearly contradicting experimental expectations. We therefore combine the results of the CI calculations with the understanding that the spins decouple to behave paramagnetically at sufficiently large impurity spacings and correspondingly negligible exchange and correlation effects.

To do so, we use a phenomenological Ansatz for the temperature dependence of the degree of spin entanglement. This is inspired by the quantum mechanical description of spin 1/2 paramagnetism, where the temperature dependence of the magnetic moment is given by $M \propto \tanh(\mu_B B/k_BT)$. For our system we replace the Zeeman term $\mu_B B$ in the argument, which describes the energy splitting of spin-up and spin-down states in a $B$ field, by the energy splitting between the two covalent CI branches $\Delta E_{\tilde b\tilde a}(s)$. The dependence of this term on the impurity spacing $s$ is shown in Fig. \ref{fig7}(a).
\begin{figure}
    \includegraphics[width=8.0 cm]{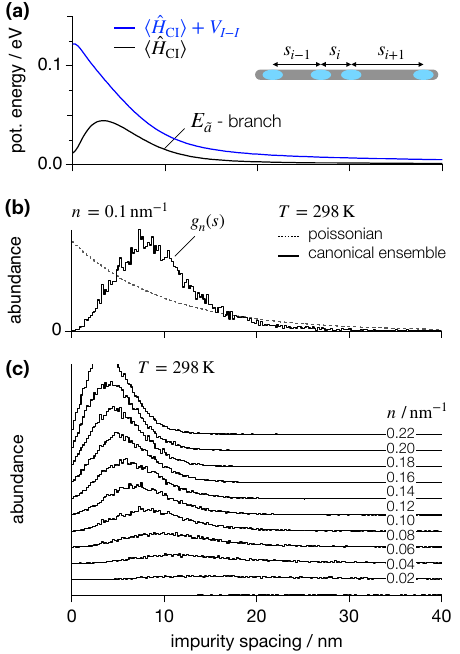}
    \caption{\textbf{Canonical distribution of impurity spacings.} (a) Impurity-impurity interaction potential with and without ion-ion interactions. (b) Monte Carlo simulation of the canonical distribution of impurity spacings at room temperature for an average impurity density of $0.1\,\rm nm^{-1}$. (c) Same as in b) but for different impurity densities.}
    \label{fig8}
\end{figure}

We can then use the $\tanh(\Delta E_{\tilde b \tilde a}/k_BT)$ energy and temperature dependence as a measure of spin entanglement. Its dependence on the impurity spacing is shown in Fig. \ref{fig7}(b). Accordingly, the dependence of the EPR signal strength of an impurity pair on its spacing $s$ is modeled by
\begin{equation}
    I_{\rm EPR}\propto \left(1-\tanh\left [\frac{\Delta E_{\tilde b \tilde a}(s)}{k_BT}\right]\right).
    \label{eq4}
\end{equation}
At a temperature of 30 K this predicts that the character of the impurity pair transitions from isolated to entangled spins at an impurity spacing of 10.4 nm. The change in impurity pair spacing required for the energy splitting $\Delta E_{\tilde b\tilde a}$ to overcome thermal energies is indicated by the vertical bars in Fig. \ref{fig7}(b).

\subsection{\label{subsec3:5}The impurity pair ensemble}

Furthermore, in order to compare the above model with experimental data, we must also take into account perturbations in the distribution of exohedrally adsorbed counterions. Such disorder leads to a certain variability of impurity pair separations which in turn will affect the degree of spin entanglement as discussed in the previous section. In the following we thus used simple Monte Carlo simulations to simulate a canonical distribution of impurity spacings in doped s-SWNTs by accounting for electronic and screened ionic interactions between impurity pairs as shown in Fig. \ref{fig8}(a).
\begin{figure}
    \includegraphics[width=8.0 cm]{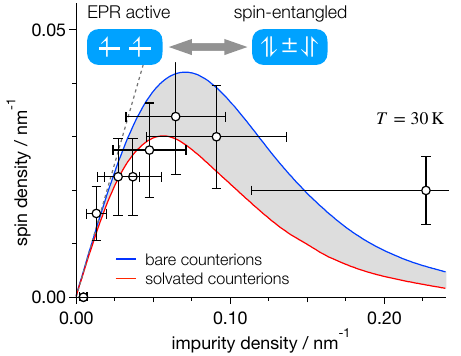}
    \caption{\textbf{Measured and simulated EPR spin signals.} Comparison of the spin density as obtained from the quantitative EPR measurements at 30 K with simulated spin entanglement, both plotted as a function of the experimentally determined impurity density. The experimental results are found to be in the range of simulated spin signals for the bare and solvated counterion scenarios. Spins become entangled at impurity densities above about $0.07\,\rm nm^{-1}$.}
    \label{fig9}
\end{figure}

In the absence of such interactions, randomly distributed impurities would give a Poissonian distribution of spacings (see the dashed line in Fig. \ref{fig8}(b). However, exchange and correlation interactions for the lowest group of CI states $E_{\tilde a}$ as well as the screened Coulomb interactions $V_{I-I}$ lead to a purely repulsive effective interaction potential between impurities (see Fig. \ref{fig8}(a)). Here the Coulomb repulsion was calculated for exohedral counterions on opposite sides of the nanotube. This predicted a canonical distribution of impurity spacings $g_n(s)$ for a mean impurity spacing $\bar s$ and associated impurity density $n=\bar s^{-1}$ as shown in Fig. \ref{fig8}(b), which differs substantially from a Poissonian distribution. Histograms obtained at room temperature from Monte Carlo simulations at different impurity densities are shown in Fig. \ref{fig8}(c). The predicted distributions show a distinct maximum at finite distances, indicating that impurities can be expected to favor uniform spacings along the nanotube, possibly leading to regularly spaced, ordered spin chains.

\begin{figure}
    \includegraphics[width=8.0 cm]{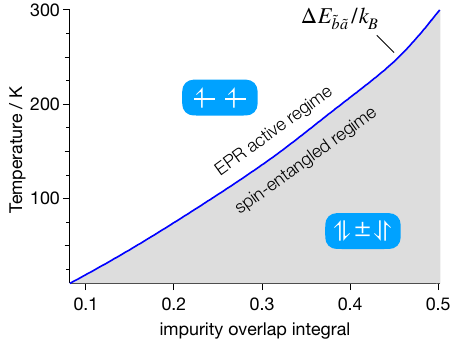}
    \caption{\textbf{EPR active and spin-entangled regimes.} The critical temperature above which thermal excitations exceed the CI energy splitting $\Delta E_{\tilde b \tilde a}$ is plotted against the impurity wavefunction overlap integral. The grey area indicates conditions where the spins of the impurity pair are expected to be entangled, while the white area indicates conditions where the spins are independent.}
    \label{fig10}
\end{figure}

\subsection{\label{subsec3:6}Comparison of modelling with EPR spin data}

To now compare the measured EPR spin signals at different doping levels $n$ with the above model, we need to weigh the degree of spin entanglement for a given impurity spacing with its abundance in the canonical ensemble: 
\begin{equation}
    \bar I_{n,\rm EPR}\propto \int_0^\infty g_n(s) \left(1-\tanh\left [\frac{\Delta E_{\tilde b \tilde a}(s)}{k_BT}\right]\right)\, d\!s.
    \label{eq4c}
\end{equation}
Doping levels $n$ are here identified with reciprocal impurity spacings. The calculations are performed as described above for impurities induced by directly adsorbed counterions as well as by solvated counterions at 30 K. As discussed earlier, impurities induced by solvated counterions have slightly smaller binding energies and slightly more extended wavefunctions.

In Fig. \ref{fig9} we compare the dependence of the calculated EPR signal strengths on the impurity density with experimentally determined spin densities. Calculations for bare (blue line) and solvated counterion induced impurities (red line) provide upper and lower brackets for experimentally measured spin densities as shown in Fig. \ref{fig9}. The modelled behaviour captures firstly the linear increase of spin density at low impurity densities (dashed line) and then secondly both the experimentally observed maximum spin density of about 0.035 (10) per nm and the position of this maximum at an impurity density of about $0.07(3)\,\rm nm^{-1}$. As expected, the maximum of the calculated spin density for solvated counterions is found at slightly lower impurity densities compared to the maximum spin density calculated for bare counterions. This is attributed to the fact that the wavefunction overlap is stronger for the solvated counterion scenario.

Finally, we can use these calculations to predict the temperature at which the transition between spin-silent, entangled spin pairs and EPR-active, independent spin regimes occurs. We plot this here as a function of the overlap integral of the impurity wavefunctions, which is a key factor in determining the formation of bonding and antibonding orbitals, as well as the magnitude of exchange and correlation effects. To illustrate this, we have plotted the critical temperature $T_c$ corresponding to the energy splitting between the two covalent branches $T_c=\Delta E_{\tilde b \tilde a}/k_B$ as a function of the impurity wavefunction overlap. In Fig. \ref{fig10} it can be seen that $T_c$ increases with wavefunction overlap. At low temperatures and high wavefunction overlap the system is entangled and spin silent, while at high temperatures and low impurity overlap the system is EPR active.

\begin{figure}
    \includegraphics[width=8.0 cm]{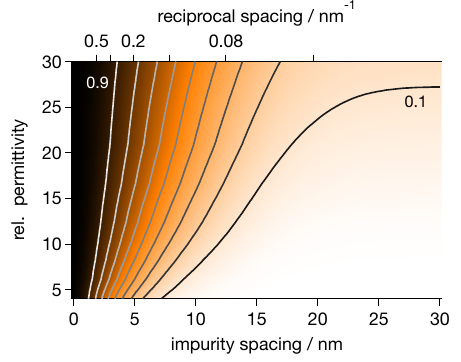}
    \caption{\textbf{Orbital overlap of impurity wavefunctions.} The orbital overlap integral of the calculated impurity wavefunctions depends on the relative permittivity as well as on the impurity spacing. Both factors are expected to determine the onset of spin-entanglement when doping levels increase.}
    \label{fig11}
\end{figure}

For completeness, in the false colour and contour plot of Fig. \ref{fig11} we have also evaluated how the orbital overlap depends on the relative permittivity used to calculate the impurity wavefunctions, as well as on the separations of the impurity pairs. This shows, perhaps somewhat counter-intuitively, that greater screening increases the sensitivity of the system to wavefunction overlap, exchange and correlation effects.

\section{\label{sec4}Conclusions}

Our study addresses an unexplored area in the field of electron paramagnetic resonance (EPR) spectroscopy of semiconducting single-wall carbon nanotubes (s-SWNTs). Previous EPR research has covered charge transfer reactions for organic photovoltaics, covalent defects, spin properties for quantum applications, and quantum information processing. However, shallow impurities introduced by redox chemical charge transfer doping -- an effective method for modifying charge transport in s-SWNTs that is critical for electronic devices -- have not been studied. Current theoretical approaches do not provide practical tools to describe the transition between non-interacting spins and spin-entangled systems. Therefore, our study also proposes a phenomenological approach to quantify spin entanglement, thus allowing a quantitative comparison of experimental and theoretical data without relying on oversimplified lattice models.

Specifically, we find that configuration-interaction (CI) calculations of interacting shallow impurity pairs in (6,5) s-SWNTs indicate that exchange and correlation effects lead to a splitting of the lower energy group of singlet and triplet spin states at impurity spacings below about 14 nm. This splitting of magnitude $\Delta E$ reflects the onset of orbital interactions as well as exchange and correlation effects, and consequently initiates spin entanglement. In our model, the degree of spin entanglement is described by a phenomenological $\tanh(\Delta E/k_BT)$ term, which allows direct comparison of the calculated energy splittings at specific impurity spacings and temperatures with experimental EPR data. It correctly describes the initial linear increase of the EPR spin 1/2 signal with doping level $n$ up to about $n=0.07\,\rm nm^{-1}$ ($T=30\,\rm K$). At this point a further increase of the doping level leads to a decrease of the spin signal due to further entanglement.

This work highlights the importance of exchange and correlation effects for the electronic structure of doped nanoscale semiconductors, where the confinement of wavefunctions and reduced screening can lead to a greater sensitivity of the electronic structure to external perturbations. Interestingly, the quasi one-dimensional arrangement of counterions on the surface of doped s-SWNTs can lead to the formation of well-ordered chains due to mutually repulsive Coulomb forces between these ions. Under suitable environmental conditions, this can lead to coherent propagation of spin waves along these nanotubes. This suggests that doped s-SWNTs could serve as a valuable platform for fundamental studies of exotic spin phenomena in one-dimensional (1D) systems, providing new insights into the behavior of spin waves and the influence of electron-electron interactions in confined geometries.

In future studies, the experimental approach might benefit from investigations with different types of redox agents that could be chosen to modify the character of the impurity states to be more or less localized. The use of divalent counterions or counterions that are more sterically distant from the nanotube could be used to do this.\cite{Murrey2023} The computational approach used here is based on a minimal basis set and could be improved in future work. In addition, we have simplified the treatment of the impurity ensemble by treating all interactions on a pairwise basis, whereas next-nearest neighbor interactions may also play a role.

With this, our study discusses a quantitative approach to compare experimental spin signal intensities with quantum mechanical calculations incorporating spin manifolds that accurately describe impurity pairs and their interactions. This marks a significant advance in our understanding of spin entanglement in doped carbon nanotubes.

\begin{acknowledgments}
We acknowledge financial support by the German National Science Foundation through the DFG GRK2112. K.E. and T.H. acknowledge additional support through grant HE 3355/4-1.
\end{acknowledgments}

\section*{AUTHOR DECLARATIONS}

\section*{Conflict of Interest}

The authors have no conflicts to disclose.

\section*{DATA AVAILABILITY}
The data supporting the results of this study are available from the corresponding author upon reasonable request.

\nocite{*}
\bibliography{literature}

\end{document}